\documentclass[twocolumn,secnumarabic,amssymb, nobibnotes, aps, prl,superscriptaddress]{revtex4-1}

\usepackage{amsmath}
\usepackage{amsfonts}
\usepackage{amssymb}
\usepackage{graphicx}
\usepackage{hyperref}
\usepackage{longtable}
\usepackage{color}

\begin{document}
\title{Hydro-osmotic instabilities in active membrane tubes}

\author{Sami C.~Al-Izzi}%
\affiliation{Department of Mathematics, University of Warwick, Coventry CV4 7AL, UK}
\affiliation{Department of Physics, University of Warwick, Coventry CV4 7AL, UK}
\affiliation{Institut Curie, PSL Research University, CNRS, Physical Chemistry Curie, F-75005, Paris, France}
\affiliation{Sorbonne Universit\'{e}s, UPMC Univ Paris 06, CNRS, UMR 168, F-75005, Paris, France}
\author{George Rowlands}%
\affiliation{Department of Physics, University of Warwick, Coventry CV4 7AL, UK}
\author{Pierre Sens}%
\affiliation{Institut Curie, PSL Research University, CNRS, Physical Chemistry Curie, F-75005, Paris, France}
\affiliation{Sorbonne Universit\'{e}s, UPMC Univ Paris 06, CNRS, UMR 168, F-75005, Paris, France}
\author{Matthew S.~Turner}%
\affiliation{Department of Physics, University of Warwick, Coventry CV4 7AL, UK}
\affiliation{Centre for Complexity Science, University of Warwick, Coventry CV4 7AL, UK}

\begin{abstract}
We study a membrane tube with unidirectional ion pumps driving an osmotic pressure difference. A pressure driven peristaltic instability is identified, qualitatively distinct from similar tension-driven Rayleigh type instabilities on membrane tubes.  We discuss how this instability could be related to the function and biogenesis of membrane bound organelles, in particular the contractile vacuole complex. The unusually long natural wavelength of this instability is in  agreement with that observed in cells.
\end{abstract}

\maketitle

The ``blueprint" for internal structures in living cells is genetically encoded but their spatio-temporal organisation ultimately rely on physical mechanisms.

A key contemporary challenge in cellular biophysics is to understand the physical self-organization and regulation of organelles \cite{Mullins2005,Chan2012}. Eukaryotic organelles bound by lipid membranes perform a variety of mechanical and chemical functions inside the cell, and range in size, construction, and complexity \cite{Alberts2008}. A quantitative understanding of how such membrane bound organelles function  have applications in bioengineering, synthetic biology and medicine. Most models of the shape regulation of membrane bound organelles invoke local driving forces, e.g. membrane proteins that alter the morphology (often curvature) \cite{Heald2014,Shibata2009,Jelercic2015}. However other mechanisms, such as osmotic pressure, could play an important role \cite{Gonzalez-Rodriguez2015}.

\begin{figure}
\includegraphics[width=0.45\textwidth,trim = 5.5cm 14.5cm 0cm 0cm,clip=true]{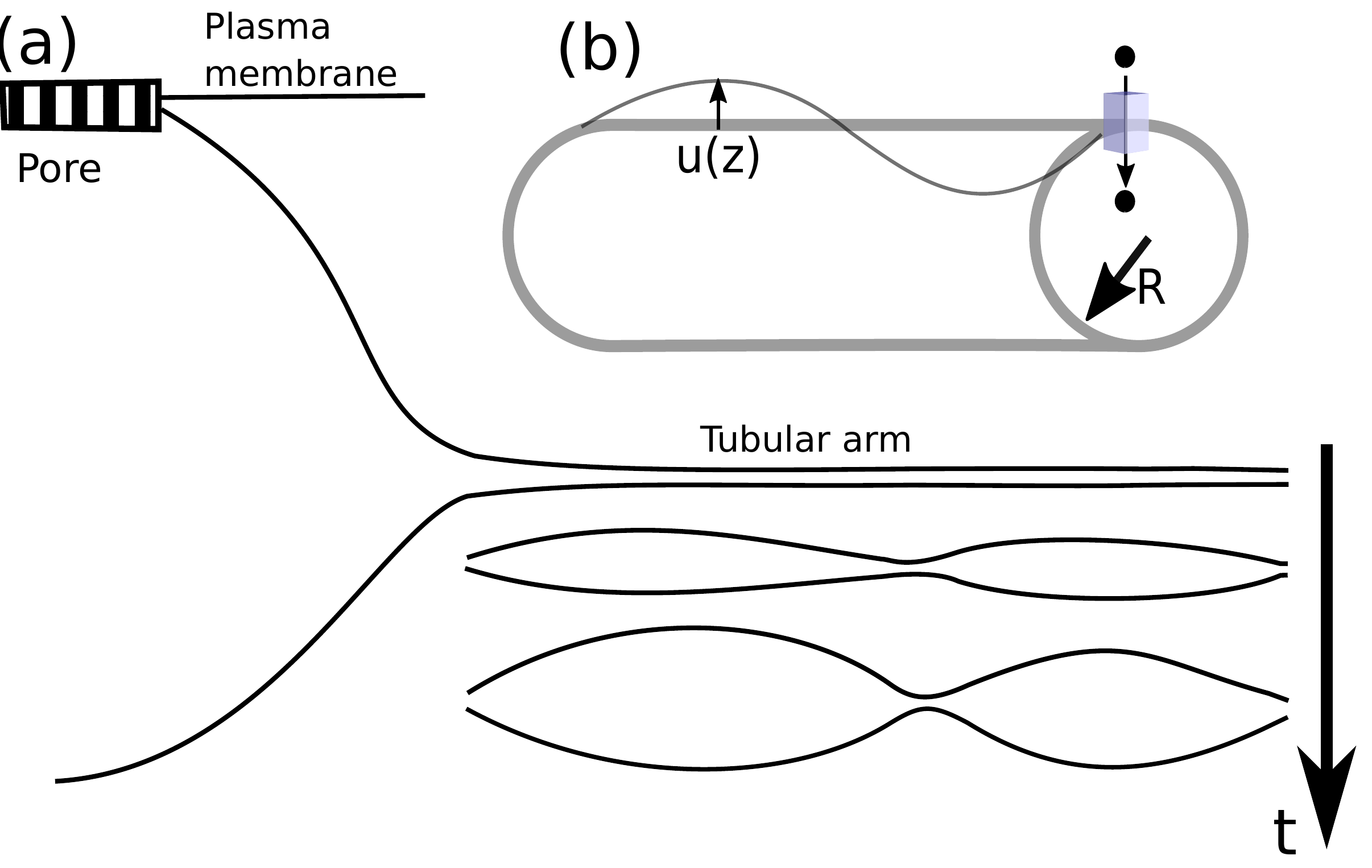}
\caption{\label{fig:CV} (a) Diagram of the contractile vacuole complex. The tube is shown connected to the main body of the CV (left). As ions are pumped in, increasing the osmotic pressure, the tube undergoes a swelling instability and undulations develop with some wavelength $\lambda$. This phenomena is observed in the contractile vacuoles of, e.g. \textit{paramecium multimicronucleatum} \cite{Patterson1980,Allen2000}. 
(b) Schematic of a membrane tube with ion pumps and surface undulations. A cartoon of a representative ion pump is shown in the top right.}\end{figure}

Membrane tubes  are ubiquitous in cells, being found in organelles such as the Golgi and endoplasmic reticulum \cite{Alberts2008} and elsewhere. Models for their formation typically involve the spontaneous curvature of membrane proteins \cite{Shibata2009} or forces arising from molecular motors, attached to the membrane, that pull tubular tethers as they move along microtubules \cite{Yamada2014}. Many of these tubules may contain trans-membrane proteins that can alter the osmotic pressure by active transport of ions. Most work on the biogenesis of cellular organelles has focused on their static morphology and generally not on their non-equilibrium dynamics. In what follows we consider an example in which the out-of-equilibrium dynamics drives
 the morphology, Fig.~\ref{fig:CV}. Our study is inspired by the biophysics of an organelle called the Contractile Vacuole Complex but additionally reveals a new class of instabilities not previously studied that are of broad, perhaps even universal, physiological relevance.

The Contractile Vacuole Complex (CVC) is an organelle found in most freshwater protists and algae that regulates osmotic pressure by expelling excess water \cite{Komsic-Buchmann2014,Stock2002,Allen2000,Naitoh1997,Docampo2013}. Its primary features is a main vesicle (CV) that is inflated by osmosis and periodically expels its contents through the opening of a large pore - probably in response to membrane tension - connecting it to the extracellular environment, thereby regulating cell volume \cite{Patterson1980,Docampo2013}. 
Water influx into the CVC is due to  an osmotic gradient generated by ATP-hydrolysing proton pumps in the membrane that move protons into the CVC \cite{Stock2002,Heuser1993,Nishi2002,Fok1995}. In many organisms such as  \textit{Paramecium multimicronucleatum}, the CVC includes several membrane tubular arms connected to the main vesicles, which are thought to be associated with the primary sites of proton pumping and water influx activity \cite{Tominaga1998}. The tubular arms do not swell homogeneously in response to water influx, but rather show large undulatory bulges with a size comparable to the size of the main CV, leading us to speculate that this might even play a role in CV formation {\it de novo}. These tubular arms appear to be undergoing a process similar to the Pearling or Rayleigh instability of a membrane tube under high tension \cite{Rayleigh1892,tomotika_instability_1935,Powers1997,Bar-Ziv1994,Bar-Ziv1997,Gurin1996,Nelson1995,Boedec2014} or an axon under osmotic shock \cite{Pullarkat2006}, but with a much longer natural wavelength: Rayleigh instabilities have a natural wave length $\lambda\sim R$ where $R$ is the tube radius. Here we derive the dynamical evolution of a membrane tube driven out-of-equilibrium by osmotic pumping.

In the CVC, the tubular arms are surrounded by a membrane structure resembling a bicontinuous phase made up of a labyrinth tubular network called the smooth spongiome (SS). We assume this to represent a reservoir of membrane keeping membrane tension constant and uniform during tube inflation.
It is possible to implement more realistic area-tension relations \citep{Boedec2014}, however this is beyond the scope of the present work.

The CVC is comprised of a phospholipid bilayer membrane. This bilayer behaves in an elastic manner \cite{Helfrich1973,Phillips2012}. At physiological temperatures these lipids are in the fluid phase \cite{Alberts2008,Phillips2012}. 
For simplicity we will treat the bilayer as a purely elastic, fluid membrane in the constant tension regime, neglecting the separate dynamics of each leaflet. 
The membrane free energy involves the mean curvature $H$ and surface tension $\gamma$  \cite{Helfrich1973,Safran2003,Nelson2008}
\begin{equation}\label{eq:freeEn}
\mathcal{F} = \int_{\mathcal{S}}\text{d}A \left(\frac{\kappa}{2}(2H)^{2} + \gamma \right)-\int \Delta P \text{d}V\text{,}
\end{equation} 
where $\text{d}A$ and $\text{d}V$ are the area and volume elements on $\mathcal{S}$, $\kappa$ is the bending rigidity, and $\Delta P$ is the pressure difference between the fluid inside and outside the tube.  Assuming radial symmetry and integrating over the volume of the tube we obtain
\begin{multline}\label{eq:cylinderEnergy}
\mathcal{F} = 2\pi\int^{\infty}_{-\infty} \text{d}z\left[\frac{\kappa}{2}r\frac{1}{\sqrt{1+\left(\partial_z r\right)^2}} \left( \frac{\partial_{zz}r}{1+\left(\partial_z r\right)^2} - \frac{1}{r}\right)^2\right.\\
 \left.+\gamma r\sqrt{1+\left(\partial_z r\right)^2} - \frac{1}{2}r^2 \Delta P\right]
\end{multline}
where $r(z,t)$ is the radial distance of the {axisymmetric} membrane from the cylindrical symmetry axis and $z$ measures the coordinate along that axis, see S.I.~for details.

We  use Eq.~(\ref{eq:cylinderEnergy}) as a model for the free energy of a radial arm of the CVC. Ion pumps create an osmotic pressure difference that drive a flux of water to permeate through the membrane. We calculate  the dominant mode of the  hydro-osmotic instability resulting from the volume increase of the tube lumen.
We write the radius of the tube as $r(z, t) = R + u(z, t)$, with $u$ assumed small, and make use of the Fourier representation $u(z,t)=\sum_q\bar{u}_q e^{\imath qz}$. Absorbing the $q=0$ mode into $R=R(t)$ allows us to write $\int u\>\text{d}z = 0$. The free-energy per unit length can be written at leading order as
\begin{equation}\label{eq:FreeEnergy}
\mathcal{F}= \mathcal{F}^{(0)} + \frac{\pi}{R}\sum_q \alpha(q)|\bar{u}_q|^2
\end{equation}
where
\begin{equation}\label{eq:alpha}
\alpha(q)=\frac{\kappa}{R^2}\left((qR)^4 -\frac{(qR)^2}{2} + 1\right) + \gamma (qR)^2 - \Delta P R
\end{equation}
and
\begin{equation}
\mathcal{F}^{(0)} = 2\pi\left(\frac{\kappa}{2 R} +\gamma R -\frac{1}{2}\Delta P R^2\right)
\end{equation}

Identifying the static pressure difference $\Delta P$ with the  Laplace pressure $P_L=-\kappa/(2R^3) +\gamma/R$, the point at which the $q=0$ mode goes unstable can be identified: the membrane tube is unstable for tube radii $R>\sqrt{3}R_{\text{eq}}$ where
$R_{eq}=\sqrt{\frac{\kappa}{2\gamma}}$ is the equilibrium radius of a tube with $\Delta P=0$.  This criterion for the {\em onset} of the instability is the same as the Rayleigh instability on a membrane tube \cite{Gurin1996}, however the instability is now driven by pressure not surface tension. This is a crucial difference. It leads to a qualitatively different evolution of the instability, as we now show. In what follows we are interested in the dynamics of the growth of unstable modes after the cylinder has reached radius $\sqrt{3}R_{\text{eq}}$.
Our initial condition is a tube under zero net pressure, although the choice of initial condition is not crucial. 
We assume that the number of proton pumps moving ions from the cytosol into the tubular arm depends only on the initial surface area, i.e.\ it is fixed as the tube volume (and surface) varies.

We denote the number of ions per unit length in the tube as $n$ and  write an equation for the growth of $n$ as
\begin{align}\label{eq:ions}
\frac{\text{d}n}{\text{d}t} = \begin{cases} 
0,\quad &t\in\left(-\infty,0\right)\\
2\pi\beta R_{eq},\quad &t\in\left[0,\infty\right)
\end{cases}
\end{align}
where $\beta$ is a constant equal to the pumping rate of a single pump multiplied by the initial area density of pumps.

The density of ions, $\rho_I$, can be obtained by solving Eq.~(\ref{eq:ions}) and dividing by volume per unit length, $v(t)$,
\begin{equation}
\rho_I = \frac{n(t)}{v(t)} = \frac{n_0}{v(t)} + \frac{2\pi\beta R_{eq} t}{v(t)}\text{.}
\end{equation}

The growth of the tube radius is driven by a difference between osmotic and Laplace pressure \cite{Rangamani2017}. This means the rate equation for the increase in volume
can be written in terms of the membrane permeability to water.
Assuming that the water permeability (number of water permeable pores) is constant during tube inflation, we write the volume permeability per unit tube length $\mu'= 2\pi R_{eq} \mu$, where $\mu$ is the (initial) permeability of the membrane.
Thus
\begin{equation}
\frac{\text{d}v}{\text{d}t} = \mu'\left(k_B T \left(\rho_I - \rho_I\left(t=0\right)\right) - P \right)
\end{equation}
where the osmotic pressure is approximated by an ideal gas law. This can be transformed into an equation for $R(t)$ on the time interval $t\in\left[0,\infty\right)$. We identify $P$ with the Laplace pressure. This leads to
\begin{equation}\label{eq:radiusRateDimensionless}
\frac{\text{d}\tilde{R}}{\text{d}\tilde{t}} = \frac{\tau_{\text{pump}}}{\tau_{\mu}}\frac{1}{\tilde{R}}\left(\frac{\tilde{t}}{\tilde{R}^2} + \left(1 + \frac{\tilde{\gamma}}{\tilde{R}}\right)\left(\frac{1}{\tilde{R}^2} - 1\right)\right)
\end{equation}
where $\tilde{\gamma}= \frac{\gamma}{k_B T R_{eq}\rho_I(t=0)}$, $\tau_{\text{pump}} = \frac{R_{eq} \rho_I(t=0)}{2\beta}$, $\tilde{t} = \frac{t}{\tau_{\text{pump}}}$, $\tilde{R} =\frac{R}{R_{eq}}$ and $\tau_{\mu} = \frac{R_{eq}}{\mu' k_B T \rho_I(t=0)}$. $\tau_{\text{pump}}$ and $\tau_{\mu}$ represent
the time-scales of pumping and permeation of water respectively. The experimental time-scale for radial arm inflation is consistent with a value of $\tau_{\text{pump}}\sim 1-10^{-1}\text{s}$. 
These dynamics assume our ensemble conserves surface tension, not volume (as in the usual Rayleigh instability). This proves to be a crucial difference.

Values of $R_{eq} = 25\text{nm}$, $\gamma= 10^{-4}\text{N m}^{-1}$ and hence $\kappa$ are estimated using experimentally measured values from \cite{Zimmerberg2006,Koster2003}. We take a typical ionic concentration in the cytosol of a protist for $\rho_I(t=0)=3.0\times10^{8}\text{$\mu$m}^{-3}$ (around $10 \text{ mMol}$) \cite{Stock2002,Phillips2012,Jackson2006}. Making an order of magnitude estimate of $\beta$ from the literature on the CVC \cite{Stock2002,Allen1988,Tani2000} leads to estimates of $\beta\sim 10^{6}\text{-}10^{9}\text{$\mu$m}^{-2}\text{s}^{-1}$. Temperature is taken as $T = 310 \text{K}$. The permeability of polyunstaurated lipid membranes is thought to be around $\mu = 10^{-4} \text{$\mu$m Pa}^{-1}\text{s}^{-1}$ \cite{Olbrich2016}. This permeability could be much larger in the presence of water channels but we find that our results are rather insensitive to increasing the value of $\mu$ because, for physiological parameter values, our model remains in the rapid permeation regime, i.e. $\tau_{\mu}/\tau_{\text{pump}}\ll 1$. This permits a multiple time-scales expansion \cite{Murray1984} of Eq.~(\ref{eq:radiusRateDimensionless}). With $\tilde{\gamma}\sim 10^{-3}\ll 1$ we find the approximate asymptotic solution
\begin{equation}\label{eq:approxSol}
\tilde{R}(t) = \left(\frac{t}{\tau_{\text{pump}}}+1\right)^{1/2} +\mathcal{O}\left(\frac{\tau_{\mu}}{\tau_{\text{pump}}}\right)\text{.}
\end{equation}
This solution agrees well with numerical solutions to Eq.~(\ref{eq:radiusRateDimensionless}). Using Eq.~(\ref{eq:approxSol}) and Eq.~(\ref{eq:alpha}) we can compute the time at which each $q$ mode goes unstable, see S.I.

We now proceed to deriving the dynamical equations for the Fourier modes. The equations governing the solvent flow are just the standard inertia free fluid equations for velocity field $\vec{v}$. These are the continuity and Stokes equations for incompressible flow
\begin{align}\label{eq:bulkCont}
\vec{\nabla}\cdot \vec{v}=0; \quad\vec{\nabla}P =\eta \nabla^2\vec{v}
\end{align}
where $P$ is the hydrodynamic pressure and $\eta =10^{-3} \text{Pa}.\text{s}$ the viscosity.
The linearised  boundary conditions are: $v_r|_{r=R}=\dot{u}+v_p$, where $v_p$ is the permeation velocity (proportional to the hydrodynamic pressure jump across the membrane:
$v_p=\mu\Delta P|_{r=R}$), and $v_z|_{r=R}=0$. The second condition is justified by invoking the membrane reservoir as a mechanism for area exchange.

Solving these equations and substituting into the membrane  force balance equation gives (in the small $qR$ limit)
\begin{equation}\label{eq:DynamicEq}
\dot{\bar{u}}_q=-\alpha_{L}(q)\left(\frac{ q^2 R(t)}{8\eta} +\frac{2\mu R(t)}{R_{\text{eq}}^3}\right)\bar{u}_q
\end{equation}
where $\bar{u}_q$ is the Fourier representation of $u$ in the $z$ direction (see S.I. for details). The response function $\alpha_L$ is obtained by replacing the static pressure difference by the Laplace pressure $P_L$ in Eq.~(\ref{eq:alpha}). Note that the term involving $\mu$, capturing mode growth due to permeation, is only relevant for wavelengths $\lambda>100R_\text{eq}$, hence we will discard it in our analysis for simplicity (but retain it in the numerics, for completeness). The growth rate for a given mode is now time dependent, hence the mode amplitude cannot be obtained from the maximum of the growth rate, but depends on the growth history and must be obtained by solving the full, time-dependent problem. We identify the instability as being fully developed when our linearised theory breaks down. 
We defined the dominant mode of the instability, called $\hat q$, as the first mode with an amplitude reaching $\sqrt{\langle|\bar{u}_{\hat q}|^2\rangle}= R_{\text{eq}}$ (a choice that does not influence our results, see S.I.). This occurs at $t=t_{\rm final}$. 

The fluctuations of modes with wavenumber $q$ about the radius $R(t)$ follow the dynamics of the Langevin equation based on Eq.~(\ref{eq:DynamicEq})
\begin{equation}
\eta(q)\dot{\bar{u}}_q = -\alpha_{L}(q)\bar{u}_{q} +\zeta_q 
\end{equation}
where $\eta(q)=\frac{8\eta}{R q^2}$ and $\zeta_q$, the thermal noise, has the following statistical properties
\begin{align}
&\langle\zeta_q\rangle = 0\\
&\langle\zeta_q(t_1)\zeta_{q'}(t_2)\rangle =\delta_{qq'}\delta\left(t_1-t_2\right) \frac{k_BT R}{\pi\eta(q)}\text{.}
\end{align}
Here the thermal noise is found using the equipartition theorem, and is thus integrated around the tube radius.

Solving this Langevin equation for $\langle |\bar{u}_q|^2\rangle$, using an initial condition of an equilibrium tube and the approximate form of $\tilde{R}(t)$ (Eq.~(\ref{eq:approxSol})) we find an integral equation for the mode growth
\begin{multline}\label{eq:FullSolution}
\frac{\langle |\bar{u}_q|^2\rangle}{R_{\text{eq}}^2} = \frac{k_BT}{2 \kappa\pi (1 + \tilde{q}^4)} e^{\left(F(0) -F(\tilde{t})\right)}\\
+ e^{-F(\tilde{t})}\int^{\tilde{t}}_{0}\frac{k_BT\tilde{q}^2 (\tilde{t}' + 1)}{\kappa\pi}\frac{\tau_{\text{pump}}}{\tau_{\eta}} e^{F(\tilde{t}')}\text{d}\tilde{t}'
\end{multline}
where $t'$ is a time variable integrating over the noise kernel (in units of $\tau_{\text{pump}}$), $\tau_{\eta}=8 R_{\text{eq}}^3\eta/\kappa$, $\tilde q=qR_{eq}$ and
\begin{multline}
F(t) = \frac{2\tau_{\text{pump}} \tilde{q}^2 \tilde{R}(t)}{15 \tau_{\eta}}\\  \times\left(40 - 5\tilde{t} + 
   \tilde{q}^2 \tilde{R}(t)^2 \left(3 \tilde{t}-2 + 6 \tilde{q}^2 \tilde{R}(t)^2\right)\right)
\end{multline}

\begin{figure}[h]
\includegraphics[width=\linewidth,trim = 3cm 9cm 4cm 9.5cm,clip=true]{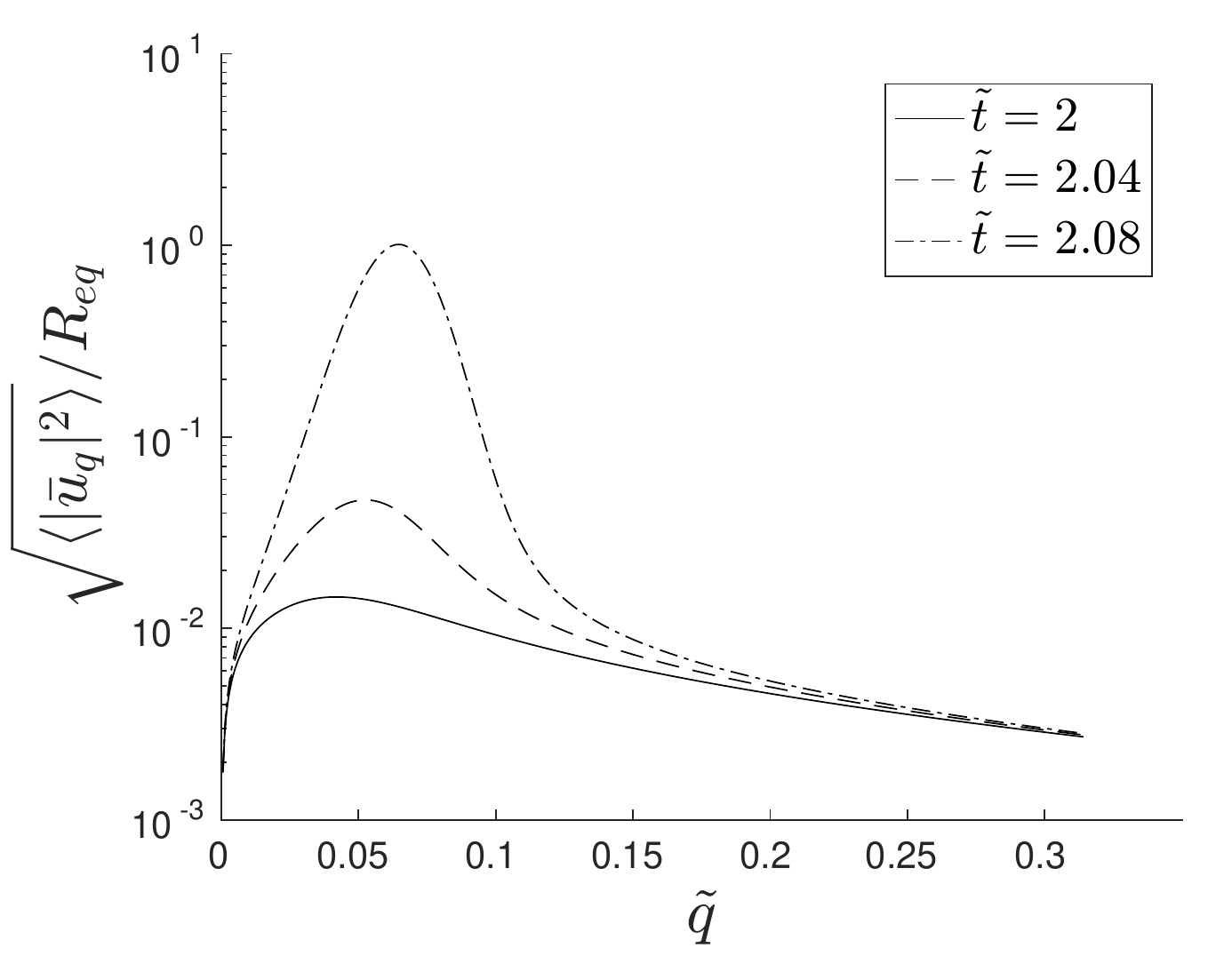}
\caption{Plot of the distibution of mode amplitude $\sqrt{\langle\bar{u}^{2}_q\rangle}$ against scaled wavenumber $\tilde{q}=qR_{\text{eq}}$ for $\tilde{t}=2.0$ (solid), $2.04$ (dashed) and $2.08$ (dash-dotted, the time when the first mode reaches $\sqrt{\langle\bar{u}^{2}_q\rangle}= R_{eq}$), $\tau_{\eta}/\tau_{\text{pump}}\sim 10^{-6}$. $R_{eq} = 25\text{nm}$, $\gamma= 10^{-4}\text{N m}^{-1}$ and $\rho_I(t=0)=3.0\times10^{8}\text{$\mu$m}^{-3}$}\label{fig:modeDist}
\end{figure}

Integrating this numerically, together with Eq.~\ref{eq:approxSol}, we can find the dynamics of the modes.
The distribution of mode amplitude against $q$ is shown in Fig.~\ref{fig:modeDist}. Although the smallest $q$ modes go unstable first, they have very slow growth  and so the mode that dominates the instability arises from the balance between going unstable early (favouring low $q$) and growing fast (favouring higher values of $q$). 

\begin{figure}[t]
\includegraphics[width=\linewidth,trim = 3cm 9.6cm 3cm 9.8cm,clip=true]{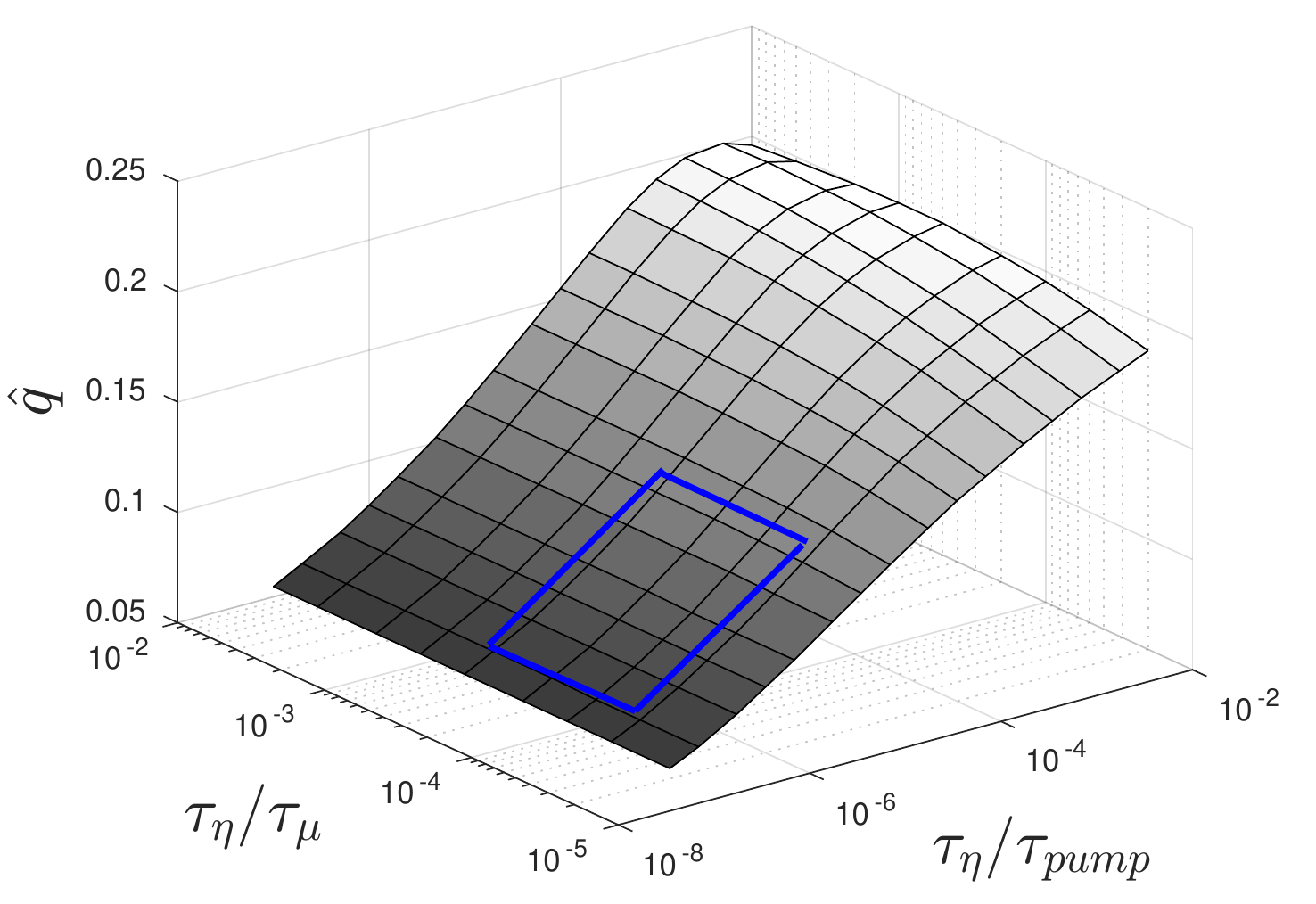}
\caption{Plot of dominant wavenumber $\hat{q}=q^{*}R_{eq}$ of the instability against ratio of viscous to pumping timescales $\tau_{\eta}/\tau_{\text{pump}}$ and ratio of viscous to permeable timescales, $\tau_{\eta}/\tau_{\mu}$. All other parameters are the same as in Fig.~\ref{fig:modeDist}. The blue rectangle indicates typical physiological parameters. }\label{fig:qVariation}
\end{figure}

We can compute numerically the natural wavelength associated with the dominant mode, $\hat q$, the first to reach $\sqrt{\langle|\bar{u}_q|^2\rangle}= R_{\text{eq}}$, see Fig.~\ref{fig:qVariation}. This gives a dominant wavelength $\lambda \sim 100\>R_{eq}\sim 2\mu\text{m}$ for parameters consistent with the CVC, much larger than that found in the Rayleigh instability, but consistent with observations of the CVC \cite{Allen2000}. 
Understanding why this is the case is not straightforward by inspection of the growth equation Eq.~\ref{eq:FullSolution}, but is more easily done by considering the time-dependent growth rate Eq.~\ref{eq:DynamicEq} (graphically presented in the S.I. - Fig.S2). 
Indeed, at the time $t=t_{\rm final}$, the dominant mode $\hat q$ whose amplitude reaches $\sqrt{\langle|\bar{u}_{\hat q}|^2\rangle}= R_{\text{eq}}$ is very close in value to the fastest growing mode (the peak of the instantaneous growth rate) at that particular time, written $q^*$, which can be derived analytically as a function of the tube radius from Eq.~(\ref{eq:DynamicEq}).
As a result of the quasi-static driving of the instability by the ion pumps, the final radius is always only marginally above the critical radius $\sqrt{3}R_{\text{eq}}$, see S.I. This is the main factor contributing to the long wavelength/small $q$ instability. In this regime, the fastest growing mode is given by $\tilde{q}^*\equiv R_{eq}q^*=\frac{1}{\sqrt{2}(3)^{1/4}}\left(\tilde{R}(t_{\text{final}})-\sqrt{3}\right)^{1/2}$  to leading order, see S.I. While a qualitatively similar regime exists for tension driven instabilities, it is only valid very close to the instability threshold and its observation would require a very precise tuning of the tension. Far from threshold, the Rayleigh or Pearling instability shows a universal relationship $\tilde{q}^{*}\sim 0.6 R_{eq}$ \cite{Bar-Ziv1997,Powers1997,Bar-Ziv1994,Boedec2014}. 
 
A related instability is that of a membrane tube under osmotic shock (see S.I.), for which one finds the most unstable mode to be $\tilde{q}^{*}\sim 0.2$. The difference between the Rayleigh and osmotic shock instabilities is due to the  growth rate having a different response when driven by a volume change compared to surface tension, see S.I. for details. The constant volume (Rayleigh) instability might be of limited relevance for the morphological changes of cellular membrane tubes, as cellular membranes typically contains a host of membrane channels, including water channels, which allow fairly rapid water transport across the membrane.  The osmotic instability that we analyse here recognises the presence of active pumps in the organelle membrane, which can drive osmotic changes in the organelle lumen \cite{Allen2000}. There is some correspondence between the fast pumping limit in Fig.~\ref{fig:qVariation} ($\tau_{\eta}/\tau_{\text{pump}}$ large) and the osmotic shock situation. The instantaneous growth rates have the same dependence in the tube radius, but have a different time dependences as the dynamics of tube inflation is different in both cases.
The osmotic shock limit is most likely not  physiologically  accessible to ion pumps. Crucially, one can see in Fig.~3 that the instability length scale  is set by {\it dynamical} parameters, most importantly the ratio of the viscosity  and pumping time-scales. 
Varying ${\tau_{\eta}}/{\tau_{\text{pump}}}$ has the effect of changing the time-scale over which the modes go unstable. It is fortuitous that the dominant wavelength does not depend strongly on the pumping rate (see S.I. - Fig.S7), the parameter we can estimate least accurately. This suggests a robustness to the wavelength selection that may have important implications for the CVC's biological function.
In the physiologically accessible range of parameters for pumping and permeation, this length scale is much larger than the asymptotic limit for either the Rayleigh instability or the osmotic shock instability.

We have developed a model for a water-permeable membrane containing uni-directional ion pumps. Hydro-osmotic instabilities realised in cells may be expected to usually lie in this class. Deriving dynamical equations for a membrane tube we identify an instability driven by this osmotic imbalance. This has a natural wavelength that is set by dynamical parameters and is significantly longer than a Rayleigh or Pearling instability but is of the same order as seen in the CVC radial arm. We speculate that this instability may provide a mechanism for biogenesis of the CV from a featureless active tube: bulges in the radial arm are similar in size to the main CV.  We will further address the question of this {\it oganellogenesis} in future work.

\acknowledgments{Thanks to J.~Prost and P.~Bassereau (Paris), M.~Polin (Warwick) and R.~G.~Morris (Bangalore) for interesting discussions and insight. Thanks to the reviewers for helpful comments and for alerting us to references \cite{Boedec2014,tomotika_instability_1935}. S.~C.~Al-Izzi would like to acknowledge funding supporting this work from EPSRC under grant number EP/L015374/1, CDT in Mathematics for Real-World Systems.}

\bibliography{Bibliography-Osmotic-instabilities}
\newpage
\onecolumngrid
\makeatletter 
\def\tagform@#1{\maketag@@@{(S\ignorespaces#1\unskip\@@italiccorr)}}
\makeatother

\makeatletter
\makeatletter \renewcommand{\fnum@figure}
{\figurename~S\thefigure}
\makeatother
\def\eq#1{{Eq.(S\ref{#1})}}    \def\fig#1{{Fig.S\ref{#1}}}
\setcounter{figure}{0} 
\setcounter{equation}{0} 
\appendix
\section{Supplementary Information}
\subsection{Differential geometry of the membrane}
This manifold has mean curvature $H$ and constant surface tension $\gamma$. The free energy for such a membrane is
\begin{equation}\label{eq:freeEn}
\mathcal{F} = \int_{\mathcal{S}}\text{d}A \left(\frac{\kappa}{2}(2H)^{2} + \gamma \right)\text{,}
\end{equation} 
where $\text{d}A$ is the area element on $\mathcal{S}$ and $\kappa$ is the bending rigidity.

For the membrane tubes in which we are interested we  parametrise the bilayer as an embedding in $\mathbb{R}^{3}$. Utilising the cylindrical symmetry of the membrane tube we write this as a surface of revolution about the $z$ axis with radius $r(z,t)$. This means that we will only consider squeezing (peristaltic) modes in our analysis. In Cartesian coordinates this surface is parametrised by the vector $\vec{R}=\left(r\cos\theta,r\sin\theta,z\right)$, i.e. by the normal cylindrical polar coordinates. From this we can induce covariant coordinates on the manifold as
\begin{align}
&\vec{e}_{1} = \frac{\partial\vec{R}}{\partial\theta} = \left(-r\sin\theta,r\cos\theta,0\right)\\
&\vec{e}_{2} = \frac{\partial\vec{R}}{\partial z} = \left(\partial_z r \cos\theta,\partial_z r \sin\theta,1\right)\text{.}
\end{align}

This allows for the definition of a Riemannian metric as
\begin{equation}
g_{ij} = \vec{e}_i\cdot\vec{e}_j \quad \text{for} \quad i,j=\{1,2\}\text{,}
\end{equation}

Hence the metric and its inverse are
\begin{equation}
g = \begin{bmatrix}
r^2 & 0\\
0 & 1+\left(\partial_z r\right)^2
\end{bmatrix}, \quad g^{-1} = \begin{bmatrix}
\frac{1}{r^2} & 0\\
0 & \frac{1}{1+\left(\partial_z r\right)^2}
\end{bmatrix}\text{.}
\end{equation}

To find the curvature of $\mathcal{S}$ we need to know how the normal vector, $\vec{n}$, to the surface $S$ varies. We can write this normal vector as 
\begin{equation}
\vec{n} = \frac{\vec{e}_1 \times\vec{e}_2}{|\vec{e}_1 \times\vec{e}_2|} = \frac{1}{\sqrt{1+\left(\partial_z r\right)^2}}\left(\cos\theta,\sin\theta,-\partial_z r\right)\text{.}
\end{equation} 

From this we can find the second fundamental form $b_{ij}=\vec{n}\cdot\vec{e}_{i,j}$ where the comma denotes a partial derivative.
Taking the determinant and trace of
\begin{equation}
b^{\text{  }j}_i = \frac{1}{\sqrt{1+\left(\partial_z r\right)^2}} \begin{bmatrix}
\frac{-1}{r} & 0\\
0 & \frac{\partial_{zz}r}{1+\left(\partial_z r\right)^2}
\end{bmatrix}\text{,}
\end{equation}
we find the mean and Gaussian curvatures
\begin{align}
&2H = \frac{1}{\sqrt{1+\left(\partial_z r\right)^2}} \left( \frac{\partial_{zz}r}{1+\left(\partial_z r\right)^2} - \frac{1}{r}\right)\\
&K = \frac{-\partial_{zz}r}{r\left(1+\left(\partial_z r\right)^2\right)^2}\text{.}
\end{align}

If we consider the case where we are only interested in membranes that do not change topology then we can apply Gauss-Bonnet theorem to the free energy term, integrating out the constant contribution of the Gaussian curvature.

\section{Dynamics of radial growth due to water permeation}
\subsection{Approximate solution for slow pumping}
\begin{figure}[h!]
\includegraphics[width=0.5\textwidth]{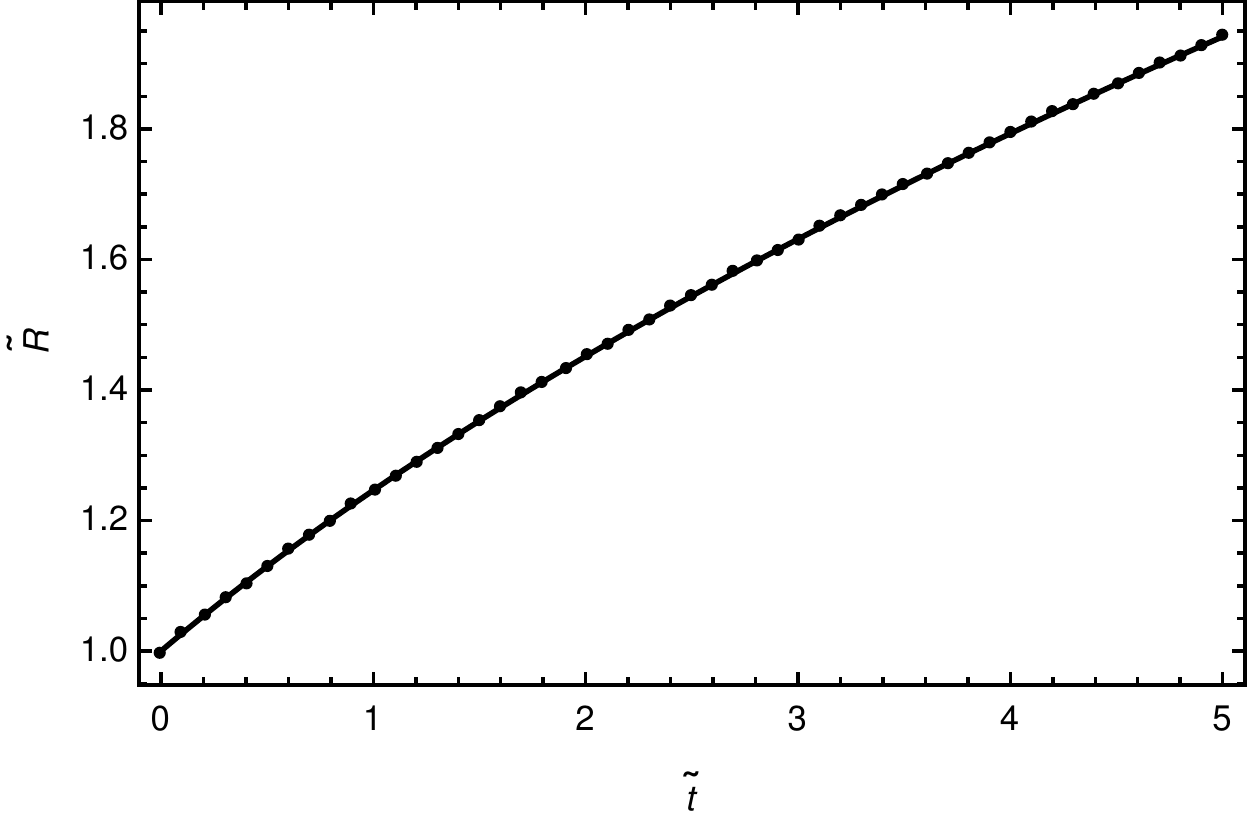}
\includegraphics[width=0.5\textwidth]{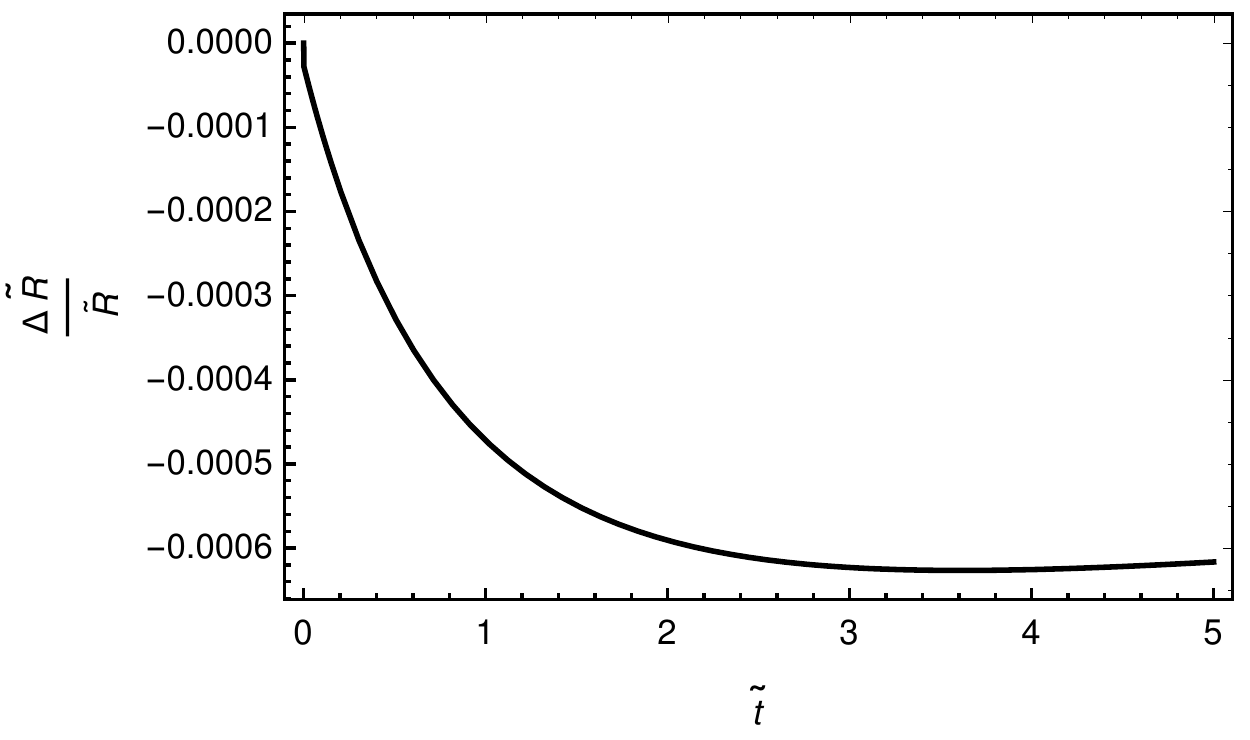}
\caption{\label{fig:approxSolError}\textbf{Left:} plot showing approximate solution (dots) and full numerical solution (solid line). \textbf{Right:} plot showing the absolute error between the approximate solution and numerical solution.}
\end{figure}
\fig{fig:approxSolError} shows the agreement between the asymptotic solution to the radial dynamics: $\tilde R(t)=\left(1+t/\tau_{\rm pump}\right)^{1/2}$ (Eq.10 - main text) and the full numerical solution of Eq.9 (main text).

We can find the radius $R(q)$ at which the mode $q$ first goes unstable by finding the zero of the $\alpha(q)$ polynomial, Eq.4 (main text), defining $R(q) = \sqrt{3}R_{eq} +\delta R(q)$. For the small $q$ limit and assuming $\frac{\delta R(q)}{R_{eq}}$ is small we find
\begin{equation}\label{eq:unstableRad}
\frac{\delta R(q)}{R_{eq}} \approx \sqrt{3} \left( R_{eq}q\right)^2 \text{.}
\end{equation}
Using \eq{eq:unstableRad} with the approximate solution for $R(t)$ gives a formula for the time the mode $q$ {\em first} goes unstable
\begin{equation}\label{eq:unstableTime}
t_q^{*} \approx \tau_{\text{pump}}\left(2+ 6\tilde{q}^2\right)
\end{equation}
where $\tilde{q}=qR_{\text{eq}}$.

\subsection{Case of an osmotic shock}
We can consider a tube with a fast-acting tension reservoir (something similar to the smooth spongiome), undergoing osmotic shock. It is interesting to understand the dominant wavelength selection in such a case as the system may be easier to implement {\it in vitro} than systems involving unidirectional ion pumps.
If the radial expansion of the membrane is driven by a hypo-osmotic shock, the radial dynamics are governed by the following growth equation
\begin{equation}
\frac{\text{d}\tilde{R}}{\text{d}\tilde{t}} = \frac{1}{\tilde{R}}\left(\frac{1}{\tilde{R}^2}\frac{\Delta \rho}{\rho_0} + \left(1 +\frac{\tilde{\gamma}}{\tilde{R}}\right)\left(\frac{1}{\tilde{R}^2}-1\right)\right)
\label{osmo_growth}
\end{equation}
where $\tilde{t} = \frac{t}{\tau_{\mu}}$, $\tilde{\gamma}= \frac{\gamma}{k_B T R_{eq}\rho_0}$, $\tilde{R} =\frac{R}{R_{eq}}$, $\tau_{\mu} = \frac{R_{eq}}{\mu k_B T \rho_0}$ and $\Delta \rho=\rho_0-\rho_{\text{shock}}$ is the change in ionic density of the outside medium due to osmotic shock. Note that the normalisation chosen here is different from the one used in the text.

\section{Solution to the stokes equations}
The equations governing the bulk flow are just the standard inertia free fluids equations for velocity field $\vec{v}$. These are the continuity equation for incompressible flow
\begin{equation}\label{eq:bulkCont}
\vec{\nabla}\cdot \vec{v}=0 
\end{equation}
and the Stokes equation
\begin{equation}
\vec{\nabla}P =\eta \nabla^2\vec{v}
\end{equation}
where $P$ is the hydrodynamic pressure and $\eta$ the viscosity.
The system has the following boundary conditions
\begin{align}\label{eq:bcs2}
&v_r|_{r=R}=\dot{u}+v_p\\
&v_z|_{r=R}=0
\end{align}
where $v_p$ is the permeation velocity, the second boundary condition is justified by any local area change in the membrane coming from exchange with the tension reservoir, the surrounding sponge phase. Hence there is no lateral membrane flow (at least at this order).

If we write the velocity field in terms of a stream function $\psi$ as
\begin{equation}
\vec{v}=\frac{1}{r}\left(\partial_z\psi\vec{e}_r -\partial_r\psi\vec{e}_z\right)
\end{equation}
the continuity equation is automatically satisfied, and the Stokes equations can be solved to give
\begin{equation}
\psi = \begin{cases}
\sum_q A_1 q r I_{1}(q r)+B_1 (qr)^2 I_0(qr) \quad r<R\\
\sum_q A_2 q r K_{1}(q r)+B_2 (qr)^2 K_0(qr) \quad r>R
\end{cases}
\end{equation}
in the interior of the tube, where $A_{1,2}$ and $B_{1,2}$ are found from the boundary conditions. $I_{\nu}(x)$ and $K_{\nu}(x)$  are modified Bessel functions of the first and second kind respectively.

From here we use the equation $v_p=\mu (\Delta P)|_{r=R}$, where $\Delta P|_{r=R}$ is the hydrodynamic pressure jump across the tube membrane, and use the solution of the interior and exterior hydrodynamic pressure from the Stokes equations to find a value of $v_p$. In Fourier space this gives
\begin{equation}
\bar{v}_{p} = \frac{\dot{\bar{u}}_{q}}{\frac{1}{2 q \eta \chi(q)\mu}-1} 
\end{equation}
where
\begin{equation*}
\chi(q) = \frac{I_0\left(I_0 - 1\right)}{qRI_0^2 -2 I_0I_1 -qR I_1^2} - \frac{K_0^2}{qRK_0^2 +2 K_0K_1 -qR K_1^2}\text{.}
\end{equation*}

The force balance equation at the membrane reads
\begin{equation}
\left(P-2\eta\partial_r v_r\right)|_{r=R} = f
\end{equation} 
where $f$ is the force required to displace the membrane 
to $u$ and can be found from the free energy. Substituting the velocity and pressure fields into this gives the dynamic equation for the modes $\bar{u}_q$
\begin{align}\label{eq:FullDynamics}
&\dot{\bar{u}}_q=-\frac{\alpha_L\left(q\right)}{2\eta R} \frac{1}{X\left(q\right)}\left(1 - 2 q \mu \chi(q)\right)\bar{u}_q\\
&X(q)=\frac{I_0\left(q R I_0-I_1\right)}{q R \left(I_1^2-I_0^2\right)+2 I_1I_0}+\frac{K_0\left(q R K_0-K_1\right)}{q R \left(K_1^2-K_0^2\right)+2 K_1K_0}
\end{align}
with the shorthand $I_{\nu}=I_{\nu}(qR)$ and $K_{\nu}=K_{\nu}(qR)$\cite{Gurin1996}. The elastic response function $\alpha_L(q)$ is obtained by replacing the pressure $P$ by the Laplace presure $P_L=\gamma/R-\kappa/(2R^3)$ in Eq.4 of the main text:
\begin{equation}
\label{alphaL}
\alpha_L(q) = \frac{\kappa}{R^2}\left((qR)^4 -\frac{1}{2}(qR)^2  +\frac{3}{2}\right) +\gamma\left((qR)^2-1\right)
\end{equation}

\eq{eq:FullDynamics} can be used to describe the dynamical instability of a membrane tube subjected to different driving mechanisms; an increase of membrane tension (Rayleigh instability), an osmotic shock, or the slow active pumping mechanism we are primarily interested in. 
In the limit $qR\ll 1$ this gives
\begin{equation}\label{eq:Dynamics}
\dot{\bar{u}}_q=-\alpha_L(q)\left(\frac{ q^2 R(t)}{8\eta} +\frac{2\mu R(t)}{R_{\text{eq}}^3}\right)\bar{u}_q
\end{equation}

\section{Mode growth rates}
\begin{figure}[t]
\center\includegraphics[width=8cm]{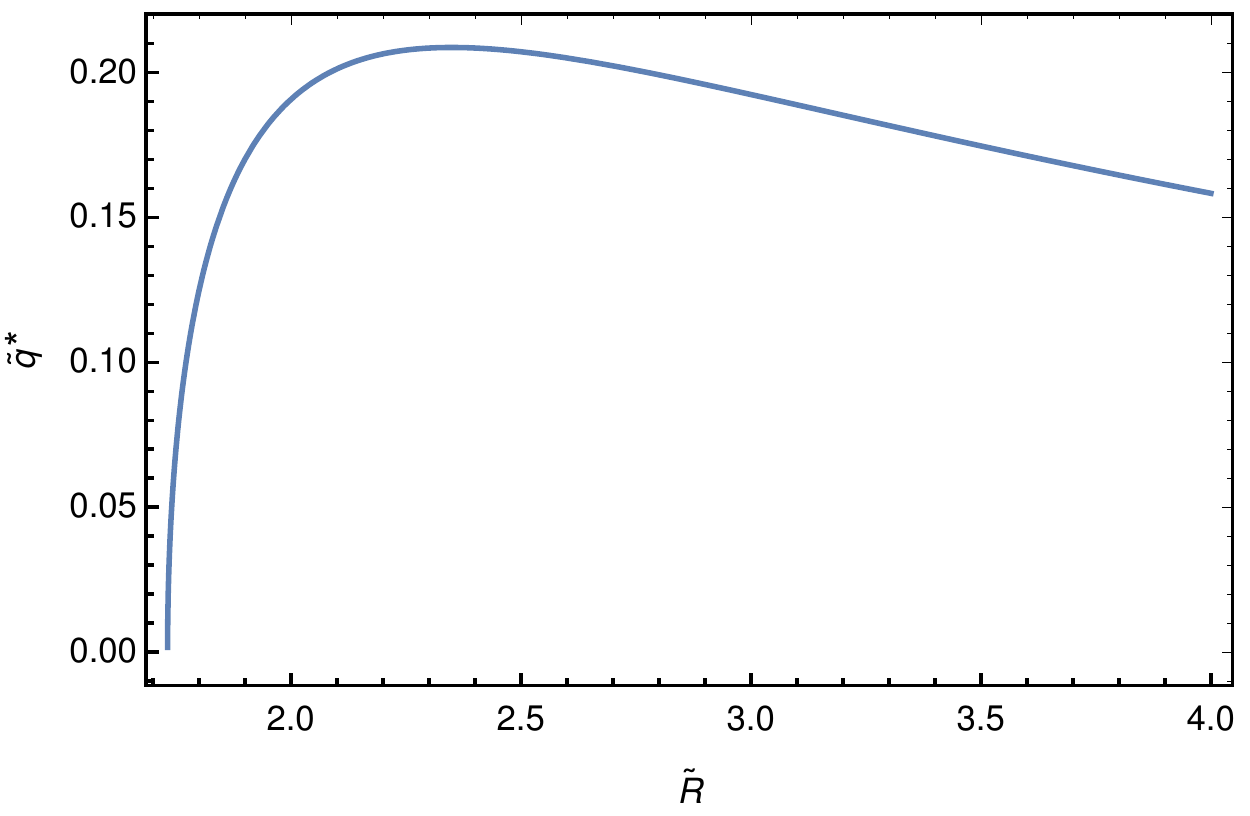}
\caption{\label{fig:maxgrowthrate} Location of the peak of the growth rate ($\tilde q^*\equiv R_{eq} q^*$) for a tube under constant tension, as a function of the tube radius. The initial tube radius $R_{eq}$ corresponds to the equilibrium radius of a tube under zero pressure.}
\end{figure}

We define the instantaneous growth rate $G(\tilde{q})=\frac{\dot{\bar{u}}_q}{{\bar{u}}_q}$ from  \eq{eq:FullDynamics}. This growth rate shows a peak as a function of $q$. The location of the peak depends on how the instability is driven. Starting with a stable tube under zero pressure with radius $R_0$ and membrane tension $\gamma_0$, the instability can be driven by an increase of tension $\gamma>\gamma^*=3 \gamma_0$ at constant volume (Rayleigh instability), or by an increase in volume (or radius) $R>R^*=\sqrt{3}R_0$ at constant tension (Osmotic instability). In the former case, and in the limit $\gamma\gg\gamma^*$, the growth rate reaches a universal shape with a peak at $R_0q^*\simeq 0.6$. The most unstable wavelength is thus entirely set by the initial tube geometry (its radius $R_0$). In the latter, the peak of the growth rate depends on the time-dependent radius and does not reach any sort of universal behaviour. In fact the location of the peak is a non-monotonic function of the radius, first increasing, then decreasing with increasing radius. Its largest  possible value is $R_0q^*\simeq 0.2$ and occurs for $R\simeq 2.35 R_0$, see \fig{fig:maxgrowthrate}.

\begin{figure}[h!]
\center\includegraphics[width=7cm]{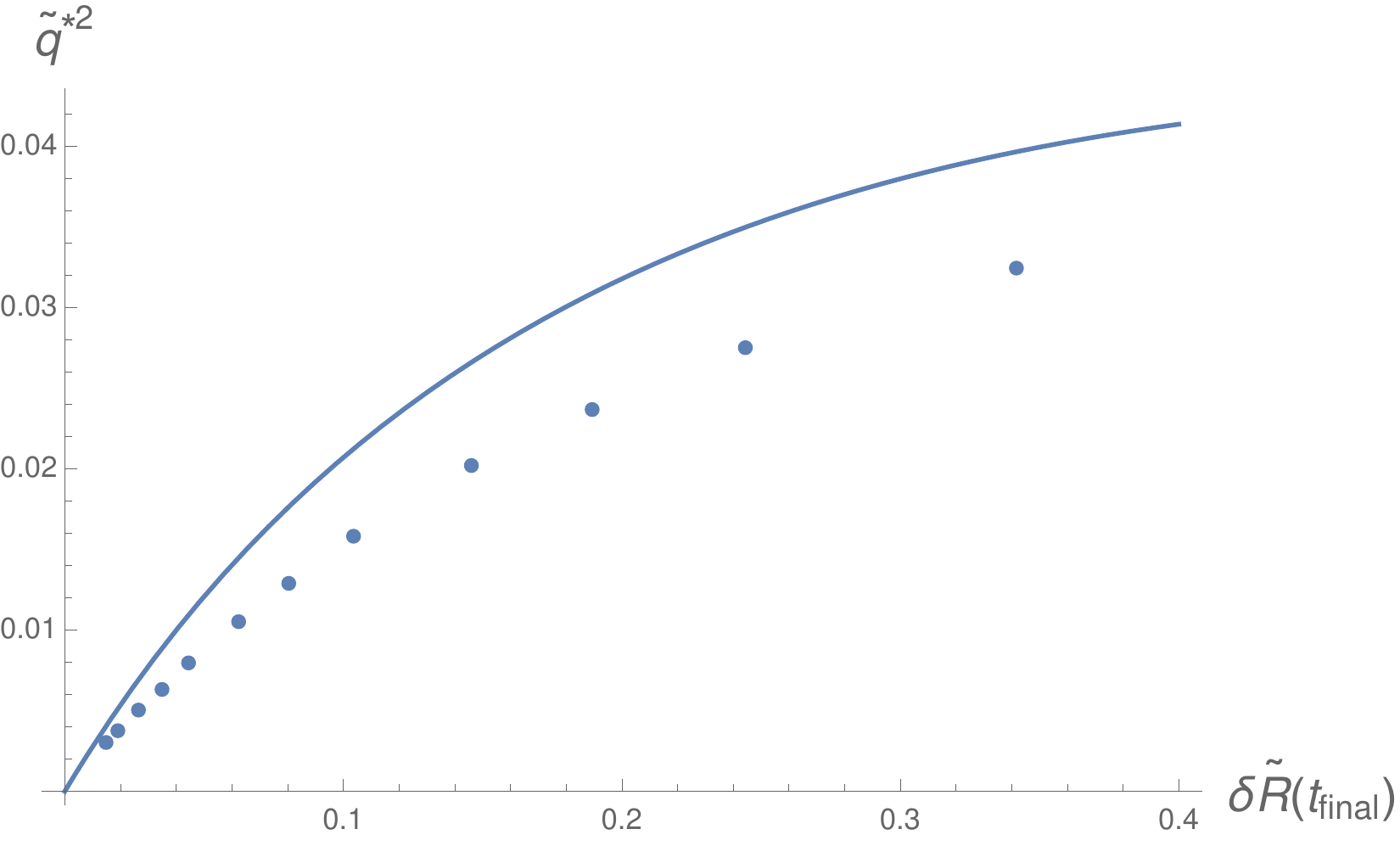}
\caption{\label{fig:qvRt}Dominant wave-number squared, $\tilde{q}^2$, plotted against final radius minus critical radius $\delta\tilde{R}(t_{\text{final}})=\tilde{R}(t_{\text{final}})-\sqrt{3}$, the solid line corresponds to the peak of the growth rate as a function of wavenumber and points represent the peak found by numerically solving the full dynamics.}
\end{figure}

As the fastest growing mode changes in time, it is the cumulative growth that is important. This means we must integrate the growth of  each $\tilde{q}$ mode over time, accounting for fluctuations, as discussed in the main text. However, due to the exponential growth the dominant $q$-mode (the one that first satisfies $\sqrt{\langle\bar{u}^{2}_q\rangle}= R_{eq}$ at $t=t_{\text{final}}$) is close to the fastest growing $q$-mode at that particular time. The latter can be expressed in terms of $\delta\tilde{R}(t_{\text{final}})=\frac{\delta R}{R_{eq}}=\tilde{R}(t_{\text{final}})-\sqrt{3}$, \fig{fig:qvRt}. It is important to note that whilst the growth rate relation does give a good approximation to the dominant wavelength, there is a difference due to the history encoded in the full dynamical description.

The peak of the growth rate relation can be found analytically ($\frac{\text{d}G}{\text{d}\tilde{q}}|_{\tilde{q}^{*}}=0$), and in the small $\tilde{q}$ limit is 
\begin{equation}\label{eq:qmax}
\tilde{q}^{*}=\frac{\sqrt{-1 - 32 \frac{\eta\mu}{R_{\text{eq}}} + 
 \frac{1}{\tilde{R}^2} + \frac{\sqrt{-17 + 4 \tilde{R}^2 (1 + 8 \frac{\eta\mu}{R_{\text{eq}}}) + \tilde{R}^4 (1 + 32 \frac{\eta\mu}{R_{\text{eq}}} (-1 + 32 \frac{\eta\mu}{R_{\text{eq}}}))}}{\tilde{R}^2}}}{\sqrt{6}}
\end{equation}
to leading order, in the $\mu\to 0$ limit, this can be expressed as $\tilde{q}^{*}=\frac{\left(\delta\tilde{R}(t_{\text{final}})\right)^{1/2}}{\sqrt{2}(3)^{1/4}}$, which is the expression given in the main text.

\begin{figure}[h]
\center\includegraphics[width=8cm]{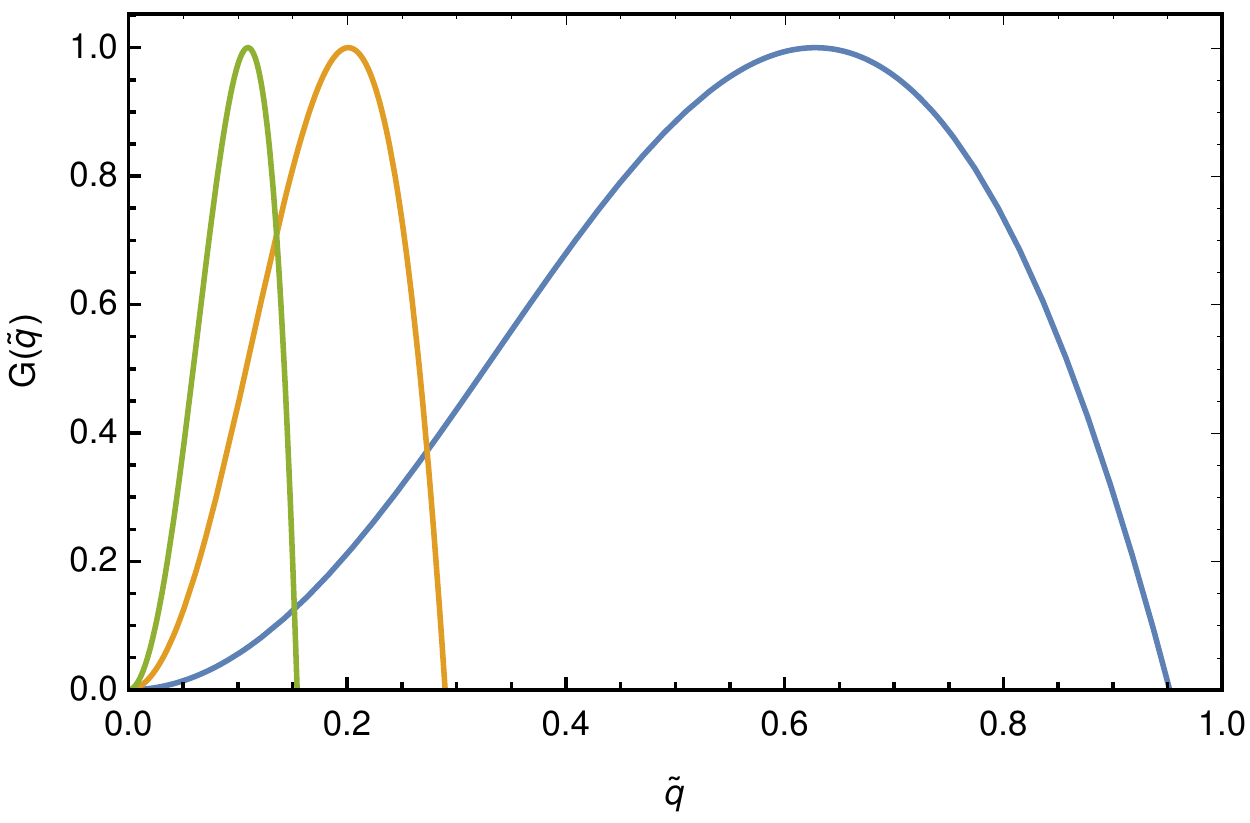}
\caption{\label{fig:Shock_Rayleigh} Normalized growth rate relation for a membrane tube undergoing a Rayleigh instability ($R_0\sim 10^{-2}\mu$m, $\mu=0$, $\kappa=10k_BT$, $\gamma=89\gamma_0$, where $\gamma_0=\kappa/(2R_0^2)$)  or responding to an osmotic shock under constant membrane tension ($R(t=0)=R_{eq}= 10^{-2}\mu$m, $\mu= 10^{-4}\mu$m Pa$^{-1}$s$^{-1}$, $\kappa=10k_BT$, $\tilde{R}(t)=2.35$). These parameters are chosen such that they illustrate the growth rate relations in the high tension limit for the Rayleigh instability (blue curve), or correspond to the maximal peak wavelength in the case of osmotic shock (orange curve). The dispersion relation for the Rayleigh instability is obtained from \eq{eq:FullDynamics}, with constant radius and the limit $\mu\to 0$. For comparison the typical growth rate for physiological parameters in the case of slow pumping (with $\tilde{R}=\sqrt{3}+0.05$) is also shown (green curve).}
\end{figure}

The growth rate relation is  quantitatively different from a Rayleigh instability due to the driving mechanism. The functional dependence of the growth rate relation depends on the polynomial $\alpha_L(q)$ describing the membrane mechanics in $q$ space (\eq{alphaL}). The Rayleigh instability  is driven by a surface tension $\gamma>\frac{3\kappa}{2R_0^2}$ at constant volume ($R(t)=R_0$), so that the magnitude of the $q^4$ term in \eq{alphaL} doesn't change. In the case of osmotic pressure however, the instability is driven by a change in volume caused by the osmotic pressure, i.e.~$R>\sqrt{\frac{3\kappa}{2\gamma}}$. This increases the prefactor to the $q^4$ term which means that the higher $q$ modes are stabilised compared to the Rayleigh case. This means that the dominant wavelength is skewed towards smaller $q$, \fig{fig:Shock_Rayleigh}.

\section{Osmotic shock}

\begin{figure}[h!]
\center\includegraphics[width=9cm, trim = 3cm 9.8cm 3cm 10cm,clip=true]{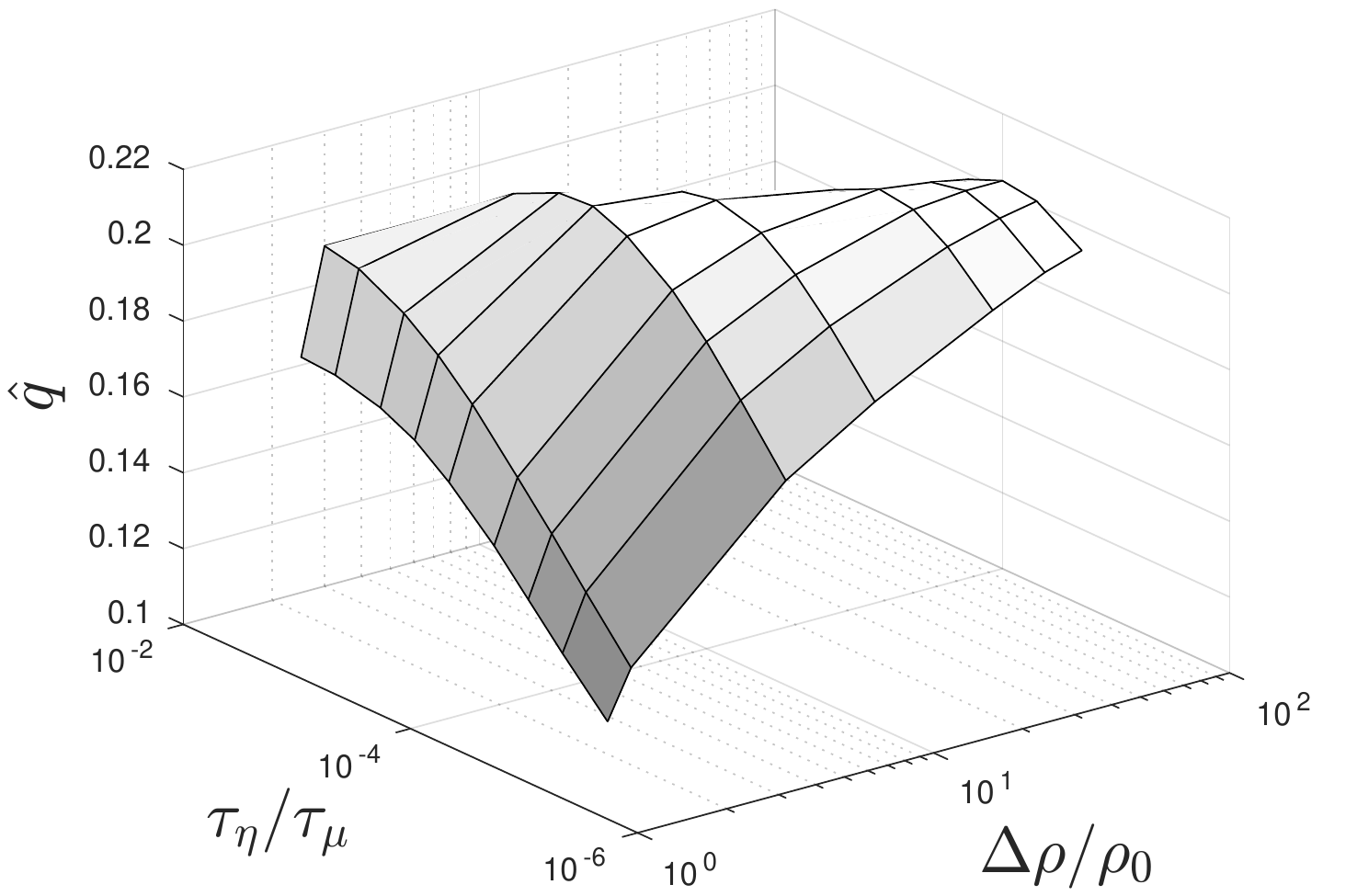}
\caption{\label{fig:OsmoticShock}Surface plot showing the dominant wave-number of an instability driven by osmotic shock when varying permeation time-scale, $\tau_\mu$, and shock magnitude $\Delta\rho/\rho_0$.}
\end{figure}

Inserting the time-dependent solution of \eq{osmo_growth} in the growth equation \eq{eq:FullDynamics} (including thermal noise, as in Eq.13 of the main text) gives access to the evolution of the amplitude of the different modes. The exact value of the dominant $\tilde{q}$ depends on the permeability $\mu$ (or the time-scale $\tau_{\mu}$) and the magnitude of the shock $\Delta\rho/\rho_0$.  A 3D plot of how this varies is shown in \fig{fig:OsmoticShock}. Comparison with the behaviour that arises in the presence of  ion pumps  (Fig.~3 - main text) shows that the peak value of the dominant mode is the same in both case, and corresponds by the peak of \fig{fig:maxgrowthrate}. This peak occurs for fast pumping ($\tau_\eta/\tau_\mu>10^{-2}$ - Fig.~3 - main text) or for strong osmotic shock ($\Delta\rho/\rho_0>10$ - \fig{fig:OsmoticShock}), showing that these two situations are somewhat similar. However the details are different due to the different dynamics of tube inflation in both cases.

The drop off in dominant wavelength of the osmotic shock instability when permeability and shock magnitude are very large is caused by the decrease of the peak of the growth rate relation at very large radii (\fig{fig:maxgrowthrate}). This happens because of a decrease in the contribution of the bending rigidity to the energy at large radii and small $\tilde{q}$. The surface tension contribution to the energy remains, hence the instability starts to be dominated by surface tension. The only contribution of the bending terms is to increasingly stabilise the larger values of $\tilde{q}$, thus pushing the peak wavelength to lower $\tilde{q}$. Interestingly the bending rigidity in this limit acts in a qualitatively similar manner to a large difference in viscosities discussed in the original fluid jet papers  \cite{Rayleigh1892,tomotika_instability_1935}.

\section{Defining the dominant wavelength}
\begin{figure}
\includegraphics[width=\textwidth,trim = 2.9cm 20cm 4cm 2cm, clip=true]{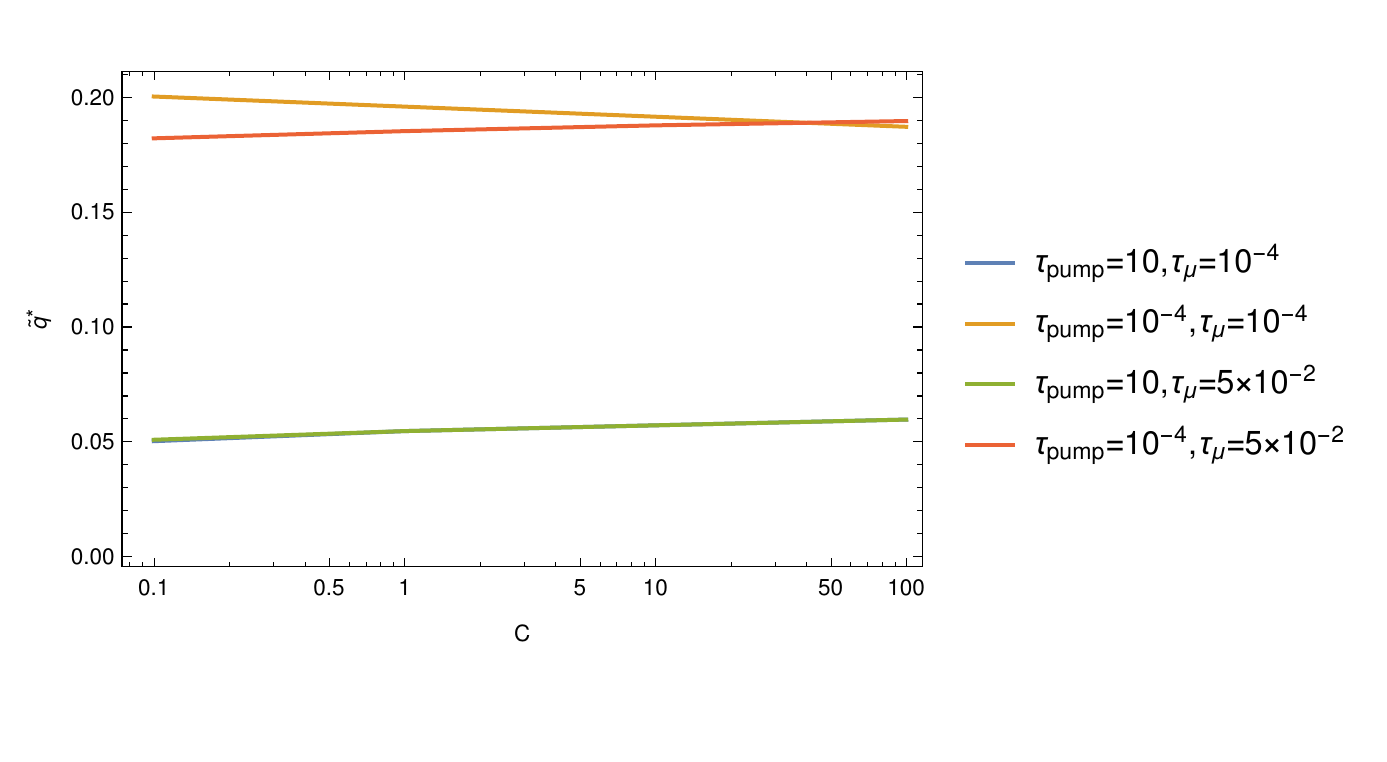}
\caption{\label{fig:cutoff}Plot of dominant wavenumber, $\hat{q}$, against cutoff criterion, $C$ (see text). Crucially any dependence on $C$ is very weak. The values of $\tau_{\text{pump}}$ and $\tau_\mu$ have been chosen to correspond with the four corners of the surface plot Fig.~3 in the main text. All times are in units of seconds.}
\end{figure}

Defining the dominante wavelength of a time dependent growth rate is in general a difficult task; as the peak of the dispersion relation is time dependent we must instead consider the full growth history of each mode. We define the dominant mode at linear order to be the first one to have $\langle |\bar{u}_q|^2\rangle=C R_{eq}^2$ where $C=1$. It is therefore a sensible thing to check that the chosen value of the cutoff, $C$, has a minimal effect on our results, i.e.~the dominant wavelength at linear order should be constant for $C\sim 1$. Plotting $\tilde{q}^{*}$ against $C$, \fig{fig:cutoff}, shows a weak logarithmic dependence of the dominant wavenumber on $C$. The only pronouced effect for a cutoff around linear ($C\sim 1$) order might be to shift the values in the fast pumping limit by $< 5\%$, the values for physiological parameters remain virtually unaffected.

\section{Weak dependence of dominant wavelength on the pumping rate in the physiological range}
\begin{figure}[h!]
\center\includegraphics[width=8cm,trim = 3cm 9cm 4cm 9cm,clip=true]{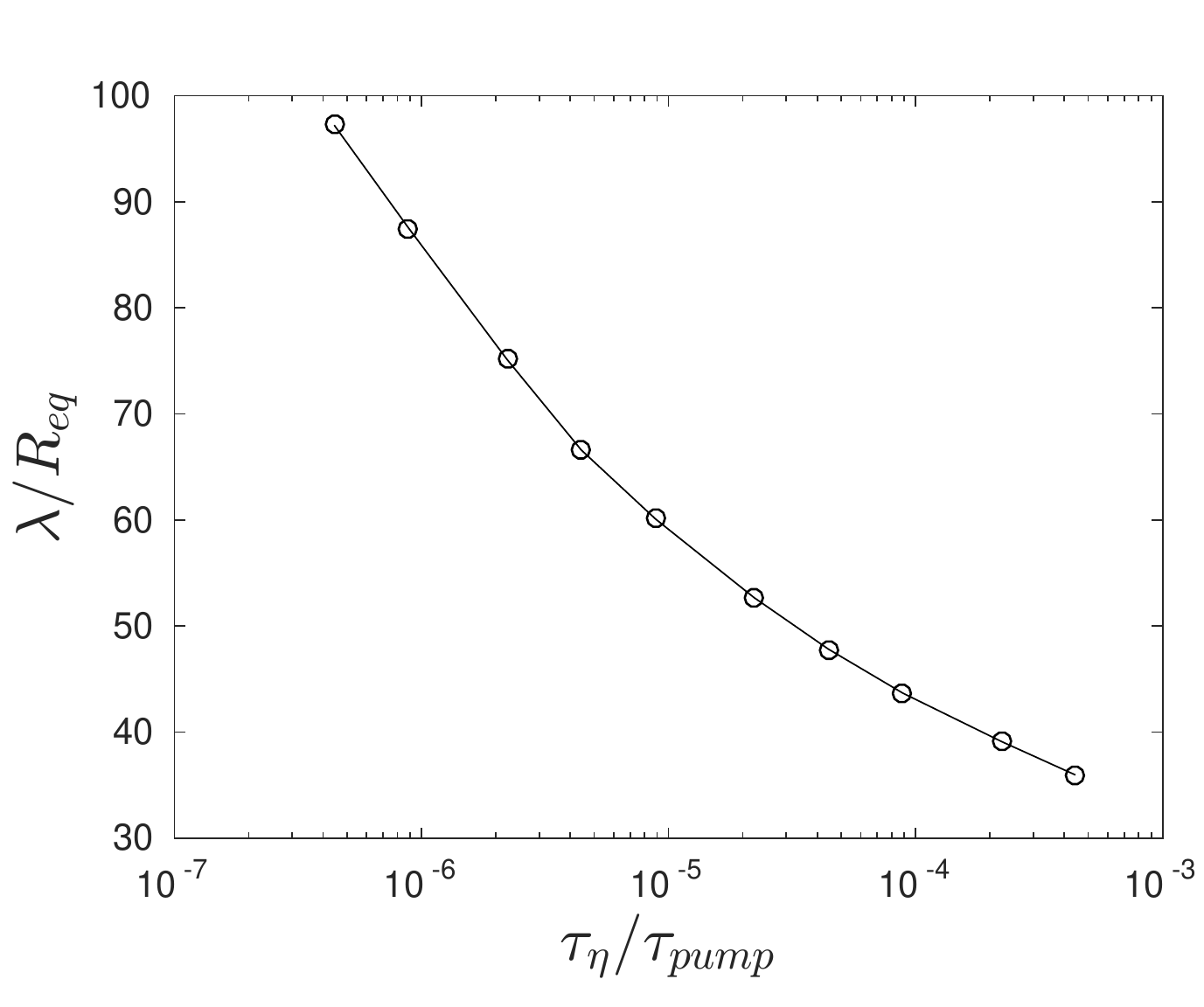}
\caption{\label{fig:tau1Variation}Plot of dominant length, $\lambda$, of instability against ratio of viscous to pumping time-scales $\tau_{\eta}/\tau_{\text{pump}}$ for the asymptotic solution found in the main paper (Eq.~10 - main text). Here $\tau_\eta/\tau_\mu=10^{-4}$. This plot is essential a cross-section of Fig.~3 in the main paper for $\tau_\eta/\tau_\mu=10^{-4}$, but plotting wavelength instead of wavenumber $\tilde{q}^{*}$.}
\end{figure}

The asymptotic solution presented in the main paper is valid for parameter estimates consistent with the CVC. It is of interest to see how the wavelength of the instability varies with pumping rate in this limit. The wavelength of the instability varies with pumping rate but very weakly (slower than logarithmically). The wavelength for time-scales consistent with the CV pumping is $\lambda\sim 1 \mu\text{m}$ which is of the correct order of magnitude for the CV and much larger than the tube radius. The weak dependence of the wavelength on the pumping provides a robust mechanism of size regulation, \fig{fig:tau1Variation}.

\section{Note on numerical implementation}
All the numerics shown in Fig.~2 and Fig.~3 of the main paper are implemented using a discrete Fourier transform, as such the autocorrolation function, $\langle|\bar{u}_q|^2\rangle$ has units of $[\text{Length}]^2$, this choice of implementation is used to simplify the criterion for the fully developed instability. The longest mode in real space is chosen to be $10^{4}R_{\text{eq}}$, this corresponds to a small enough spacing for the $q$ space to approximate a continuum.

\end{document}